\documentclass[prb,twocolumn,amsmath,amssymb,showpacs]{revtex4}
\usepackage{graphicx}
\usepackage{amssymb}

\def\beq{\begin{eqnarray}}
\def\eeq{\end{eqnarray}}
\def\beqa{\begin{eqnarray}}
\def\eeqa{\end{eqnarray}}

\begin{document}

\title{Self-energy effects in cuprates and the dome-shaped behavior of the superconducting critical temperature}

\author{Guillermo Buzon, Adriana Foussats, Mat\'{i}as Bejas, and Andr\'es Greco}
\affiliation{Facultad de Ciencias Exactas, Ingenier\'{\i}a y Agrimensura and
Instituto de F\'{\i}sica Rosario
(UNR-CONICET).
Av. Pellegrini 250-2000 Rosario-Argentina
}

\date{\today}

\begin{abstract}

Hole doped cuprates show a superconducting critical temperature $T_c$  
which follows an universal  dome-shaped behavior as function of doping.
It is believed that the origin of superconductivity
in cuprates is entangled with the physics of the pseudogap phase. 
An open discussion is whether the source of superconductivity is the same that causes the pseudogap properties. 
The $t$-$J$ model treated in large-N expansion shows
$d$-wave superconductivity  triggered by non-retarded interactions, and
an  instability of the paramagnetic state to a flux phase or $d$-wave charge density wave ($d$-CDW) state.  
In this paper we show that self-energy effects near $d$-CDW instability may lead   
to a dome-shaped behavior of $T_c$.
In addition, it is also shown that these self-energy contributions may
describe several properties observed in the 
pseudogap phase. In this picture, although fluctuations responsible for the pseudogap properties leads
to a dome-shaped behavior, they are not involved in pairing which is mainly non-retarded.

\end{abstract}

\pacs{74.72.-h, 74.72.Gh, 74.72.Kf, 74.25.Dw}

\maketitle

\section{Introduction}

In spite of many years of research the origin of superconductivity in cuprates is still 
controversial. The superconducting critical
temperature ($T_c$)  follows an universal dome-shaped behavior (DSB) 
where its maximum value takes place at optimal doping ($\delta \sim 0.16$), which separates underdoped (UD) 
from overdoped (OD) regions.
Another universal feature is the $d$-wave symmetry of the
superconducting gap which has a maximum on the Fermi surface (FS) point in the
$(0,0)$-$(0,\pi)$ direction 
(antinodal direction) 
and  vanishes upon approaching the Brillouin zone diagonal (nodal direction).\cite{timusk99} 
In addition, UD and OD regions show very 
distinct normal state properties as observed, for instance, by angle-resolved photoemission spectroscopy (ARPES) 
and transport  experiments.   Several macroscopic and microscopic experiments
support the point of view that in UD  a pseudogap (PG) appears in the normal state at a temperature
$T^*$ which is well above $T_{c}$ and, in contrast to $T_{c}$,  
increases with decreasing doping.\cite{timusk99}

In spite of the consensus about the existence of the PG, its origin and
nature remain elusive. 
Two major scenarios were proposed to explain the PG. 
In one scenario it originates from
preformed pairs above $T_{c}$.\cite{emery95,norman07,chien09}  
In the second one, the PG is distinct from the
superconducting gap and associated with a certain order which competes 
with superconductivity.\cite{vishik10,kondo11,yoshida12} 
Several phenomenological models for the two-gap scenario 
have been proposed, but invoking different orders:  
$d$-wave charge density wave ($d$-CDW),\cite{chakravarty01} 
$d$-wave Fermi surface deformations,\cite{yamase09}  
charge density wave\cite{castellani95,becca96,hashimoto10,he11} 
including stripes,\cite{kivelson03,vojta09} 
phase separation,\cite{emery93,castellani95,becca96} 
and others such as 
resonating-valence-bond-type charge order\cite{yang06}  
and loop-current order.\cite{varma06,varma99} 

One dispute about the origin of superconductivity is whether pairing
is originated by  non-retarded  or retarded interactions.\cite{maier08}
In the latter case, as in the electron-phonon (e-ph) conventional superconductivity, 
the excitations responsible for the pairing glue should be also present in transport and dynamical properties 
in the normal state (see Ref.[\onlinecite{vanheumen09}] and references therein). 
On the other side, in the former case transport and
dynamical properties in the normal state are not affected by the pairing interaction.\cite{greco01}
Non-retarded interactions were proposed earlier\cite{anderson87} and 
 still firmly supported.\cite{anderson07} Recent experiments 
 suggest that pairing may be certainly non-retarded.\cite{mansart13,park13}

The common belief is that superconductivity is entangled with
the PG physics. Thus, a model candidate for describing cuprates must show $d$-wave superconductivity following a DSB and 
its relation with properties observed in the PG phase. 
The $t$-$J$ model treated in large-N expansions shows, at low doping and low temperature, 
tendencies to several kinds of instabilities, being the flux phase (FP) the leading one.\cite{affleck88,cappelluti99,bejas12}
Below a transition temperature $T_{FP}$ the FP is developed and the 
translational symmetry broken.
The FP can be associated to the  $d$-CDW phase which 
was proposed phenomenologically for describing
the PG phase in cuprates.\cite{chakravarty01} $d$-CDW and FP are characterized 
by orbital currents flowing in a staggered pattern around each
plaquette.
Although the FP bears several characteristics for being considered as a candidate 
for the PG, the breaking of the translational symmetry and the existence of a phase transition 
are not clear from the experiments. On the other hand, 
non-retarded $d$-wave superconductivity triggered by the antiferromagnetic exchange $J$ is also observed
in the $t$-$J$ model but
with $T_c$ that monotonically increases with decreasing doping.\cite{cappelluti99,zeyher98}

The path integral large-N approach for the $t$-$J$ model in the context of 
Hubbard operators,\cite{foussats04} allows the computation of fluctuations above the mean-field level
and the calculation of self-energy effects.\cite{bejas06}
These self-energy effects were recently
confronted with experiments in the normal state, such as angle-dependent 
magnetoresistance (ADMR)\cite{buzon10}
and ARPES,\cite{greco09,greco08,bejas11,greco11,foussats08} after which they show qualitative agreement. We have shown
that several aspects related to the Fermi arc phenomenology and ADMR features
can be described considering self-energy effects near the FP instability.

In this paper we discuss the possibility that in the proximity to the FP instability, 
self-energy effects and non-retarded paring play together and develop a DSB for $T_c$. At the same time 
this process may describe several normal state properties observed in
the pseudogap phase.

The paper is organized as follows. In Sec. \ref{BG} we review the background supporting 
present calculation. 
The FP instability of the paramagnetic
phase, the existence of a quantum critical point (QCP), and the main characteristics of the dynamical 
flux susceptibility are
presented.
In Sec. \ref{FAMF} self-energy effects in the proximity to the flux instability, 
their  role on normal state properties as those observed in 
ARPES and ADMR, and their  importance for describing the pseudogap features are discussed. In Sec. \ref{flux} 
the interplay between SC and the self-energy associated with the PG is discussed. 
Discussion and conclusion are presented in Sec. \ref{Conclusion}. 

\section{Background}\label{BG}

The $t$-$J$ model, derived from the Hubbard model 
in the strong coupling limit,\cite{chao77}  is considered one of the basic models for studying 
correlated systems as high-$T_c$ cuprates. The model is described by the Hamiltonian

\begin{equation}
H = -\sum_{i, j,\sigma} t_{i\,j}\tilde{c}^\dag_{i\sigma}
\tilde{c}_{j\sigma}
+ J \sum_{<i, j>} (\vec{S}_i \cdot \vec{S}_j-\frac{1}{4} n_i n_j)
\end{equation}

\noindent where $t_{i\,j} = t$ $(t')$ is the hopping integral between the 
first (second) nearest-neighbor sites on a square lattice,  
$J$ is the Heisenberg exchange interaction between nearest-neighbor sites, 
$n_i$ is the electron density operator and 
$\vec{S}_i$ is the spin. 
$\tilde{c}^\dag_{i\sigma}$ and $\tilde{c}_{i\sigma}$ are 
the creation and annihilation operators of electrons 
with spin $\sigma$ ($\sigma=\downarrow$,$\uparrow$) respectively,
under the constraint that double occupancy of electrons is excluded 
at any site $i$. 

Although the simple appearance of this model, $\tilde{c}^\dag_{i\sigma}$ and $\tilde{c}_{i\sigma}$ 
are Hubbard operators which makes the treatment 
highly nontrivial.\cite{ovchi} 
Several numerical methods have been applied for studying this model. Among others stand: the Quantum Monte-Carlo, 
which is suitable for calculating spectral functions for one-hole case,\cite{brunner00} while for finite
doping the problem  of the sign makes the calculation uncontrollable; and the Lanczos diagonalization\cite{dagotto94}  and its finite 
temperature version\cite{jaklic00} which are limited to finite clusters.
Among the analytical methods one of the most popular is the slave-boson approximation where $\tilde{c}^\dag_{i\sigma}$ and $\tilde{c}_{i\sigma}$
are decoupled in terms of fictitious particles.\cite{wang92} 
However, the decoupling scheme introduces a gauge degree of freedom which requires gauge 
fixing and complicates the treatment of fluctuations above mean-field level. \cite{wang92,kotliar91}
See Ref. [\onlinecite{shastry11}] for recent discussion about the 
interest in analytical computation of self-energy fluctuations.

In order to avoid these difficulties, a large-N expansion for the $t$-$J$ model was formulated in terms of 
a path integral for Hubbard operators. In this formulation $\tilde{c}^\dag_{i\sigma}$ and $\tilde{c}_{i\sigma}$
are treated without invoking fictitious particles. 
In the large-N approach the two spin components are extended to N
and an expansion in powers of the small parameter 1/N is performed.
In spite that the physical spin projection is two and not N, the
large-N expansion provides a controllable
framework with no perturbative expansions in any model parameter, by which the
results are in strong coupling. On the other hand, results obtained
in large-N are improved with increasing doping. \cite{bejas06}
This large-$N$ treatment yields, in the square lattice, 
to a paramagnetic state with a quasiparticle (QP) dispersion 
\begin{eqnarray}
\epsilon_{{\bf k}}=&-&(t \delta+rJ)[\cos(k_x)+\cos(k_y)]\\ \nonumber
&+&2t' \delta \cos(k_x) \cos(k_y)-\mu\;,
\end{eqnarray}
where the contribution  
$r$ to the mean-field band and the chemical potential $\mu$ must be obtained
self-consistently\cite{foussats04} from  

\begin{eqnarray}
r=\frac{1}{N_s} \sum_{\bf k} \cos(k_x) n_F(\epsilon_{{\bf k}})\;,
\end{eqnarray}
and

\begin{eqnarray}
1-\delta=\frac{2}{N_s} \sum_{\bf k} n_F(\epsilon_{{\bf k}})\;,
\end{eqnarray}
where $n_F$ is the Fermi function, $\delta$ is the doping away from half-filling, and
$N_s$ is the number of sites.

The path integral large-N expansion allows to go beyond mean-field and  
to compute self-energy fluctuations in ${\cal O}(1/N)$.  
The expression for the self-energy is \cite{bejas06}

\begin{eqnarray}\label{SigmaIm}
 {Im}\Sigma({\bf k},\omega) = -\frac{1}{N_{s}}
\sum_{{\bf q},a,b} h_{a}({\bf k},{\bf q},\omega-\epsilon_{{{\bf k}-{\bf q}}}) h_{b}({\bf k},{\bf q},\omega-\epsilon_{{{\bf k}-{\bf q}}}) \nonumber \\ 
\times {Im}[D_{ab}({\bf q},\omega-\epsilon_{{{\bf k}-{\bf q}}})]
 [n_{F}( -\epsilon_{{\bf k}-{\bf q}}) +n_{B}(\omega-\epsilon_{{{\bf k}-{\bf q}}})] \nonumber \\
\end{eqnarray}
\noindent where $n_B$ is the Bose factor, and the vector $h_{a} ({\bf k},{\bf q},\nu)$ is
\begin{widetext}
\begin{eqnarray}
h_{a} ({\bf k},{\bf q},\nu)
&=&\left\{ \frac{}{} \right. \frac{2\epsilon_{{\bf k}-{\bf q}}+\nu+2\mu}{2}
  + J r \left[ \cos\left(k_x-\frac{q_x}{2} \right) \cos \left( \frac{q_x}{2} \right) +
\cos\left(k_y-\frac{q_y}{2} \right) \cos \left( \frac{q_y}{2} \right) \right]
 \; ; 1 \; ;\nonumber \\
&& -J r \; \cos \left( k_{x}-\frac{q_{x}}{2} \right) \; ;
-J r \; \cos \left( k_{y}-\frac{q_{y}}{2} \right) \; ; \; 
 J r \; \sin \left( k_{x}-\frac{q_{x}}{2} \right) \; ; \;
J r \; \sin \left( k_{y}-\frac{q_{y}}{2} \right) \left. \frac{}{} \right\}.
\end{eqnarray}
\end{widetext}

The information contained in the vector $h_{a} ({\bf k},{\bf q},\nu)$ is the following: 
the first component corresponds to the usual charge channel, the second component 
corresponds to the non-double occupancy constraint, and the last four correspond to the Heisenberg
exchange channels, thus, these exists only for finite $J$.

In Eq.(\ref{SigmaIm}) $D_{ab}$ is a $6\times6$ matrix which contains contributions from 
the six different channels and their mixing. $D_{ab}$ describes all possible types of charge 
susceptibilities\cite{bejas12} and can be written as

\begin{equation}
D^{-1}_{ab}({\bf q},i\nu_{n})=[D^{(0)}_{ab}({\bf q},i\nu_{n})]^{-1}- \Pi_{ab}({\bf q},i\nu_{n})
\end{equation}
\noindent where

\begin{widetext}
\begin{equation} \label{eq:D0-1}
D^{(0)}_{ab}({\bf q},i\nu_{n}) = 
\left(
  \begin{array}{cccccc}
 \delta^2/2 (V-J/2) [\cos(q_x)+\cos(q_y)]      & \delta/2   & 0         & 0       & 0     & 0 \\
   \delta/2     & 0            & 0                         & 0       & 0     & 0 \\
  0            & 0            & J\;r^2                   & 0       & 0     & 0 \\
  0            & 0            & 0                         & J\;r^2   & 0     & 0 \\
  0            & 0            & 0                         & 0         & J\;r^2   & 0   \\
  0            & 0            & 0                         & 0          & 0     & J\;r^2  \\
 \end{array}
\right)^{-1}
\end{equation}
\noindent and
\begin{equation}
\Pi_{ab}({\bf q}, i\nu_{n})
=- \frac{1}{N_s} \; \sum_{{\bf k}}\;h_{a}({\bf k},{\bf q},\epsilon_{\bf k}-\epsilon_{{\bf k}-{\bf q}}) \;h_{b}({\bf k},{\bf q},\epsilon_{\bf k}-\epsilon_{{\bf k}-{\bf q}})
\; g({\bf k},{\bf q},i\nu_{n}) 
- \delta_{a}^R\delta_{b}^R\;\frac{1}{N_s} \sum_{{\bf k}} \frac{\epsilon_{{\bf k}-{\bf q}} - \epsilon_{{\bf k}}}{2} \;
n_{F}(\epsilon_{{\bf k}}) \; , 
\end{equation}
\end{widetext}

\noindent with 
\begin{eqnarray} \label{eq:gPi}
g({\bf k},{\bf q},i\nu_{n}) = \frac{[n_{F}(\epsilon_{{\bf k} - {\bf q}}) - n_{F}(\epsilon_{{\bf k}})]} 
{i\nu_{n} + \epsilon_{{\bf k} - {\bf q}} - \epsilon_{{\bf k}} } \; ,
\end{eqnarray}
\noindent where $i\nu_n$ are bosonic Matsubara frequencies.

The instability of the paramagnetic phase is indicated by the divergence of the 
static susceptibilities defined by $D_{ab}({\bf q},i\nu_n=0)$.\cite{foussats04}
As in Ref.[\onlinecite{bejas12}] we study eigenvalues and eigenvectors 
of the $6 \times 6$ matrix $[D_{ab}({\bf q}, i\nu_n)]^{-1}$ at $i\nu_n=0$. 
When an eigenvalue crosses zero at a given doping rate, temperature $T$ 
and momentum $\bf q$, then an instability occurs toward a phase characterized 
by the corresponding eigenvector \textbf{\textit{v}}.

\subsection*{Flux phase instability of the paramagnetic state}\label{BG_A}

Here we choose suitable parameters for cuprates $t'/t=0.35$, $J/t=0.3$, where the realistic value 
for $t$ is around $400 \, meV$. In what follows the 
lattice constant $a$ and the hopping $t$ are 
considered as length and energy units, respectively.

At $T=0$ the leading instability
is the flux phase which occurs at 
$\delta=\delta_c\sim 0.17$ for ${\bf q}\sim {\bf Q}=(\pi,\pi)$, while at finite $T$ the instability occurs at ${\bf q}={\bf Q}$.
The flux phase instability is associated with an eigenvector $\textbf{\textit{v}}\sim \frac{1}{\sqrt{2}}(0,0,0,0,1,-1)$ of the 
$D_{ab}({\bf q},i\omega_n=0)$ susceptibility matrix, 
which leads to staggered circulating currents and a flux that penetrates the
plaquette in the square lattice.\cite{cappelluti99,affleck88,bejas12} 

At this point is worth to note that similar circulating currents 
where proposed by Varma for describing 
the PG.\cite{varma97} However, this picture consists in counter circulating currents within the $CuO_2$ unit cell and do not break 
the translational symmetry as the flux phase. 

Below the transition temperature $T_{FP}$, the flux phase is characterized by a $2 \times 2$ Green function 

\begin{equation} \label{G0phi}
\widehat{G}^{-1}({\bf k},i\omega_{n}) = 
\left(
  \begin{array}{cccccc}
 i\omega_n-\epsilon_{\bf k}  &i \Phi \gamma({\bf k}) \\
 -i\Phi \gamma({\bf k})     &i\omega_n-\epsilon_{\bf k-Q}\\
\end{array}
\right).
\end{equation}

In Eq.(\ref{G0phi})
$i\omega_n$ are fermionic Matsubara frequencies and $\gamma({\bf k})=\frac{1}{2}(\cos k_x-\cos k_y)$, 
showing the $d$-wave character of the FP 
order parameter $\Phi$. This order parameter satisfies the following equation

\begin{eqnarray}\label{phi}
\Phi=-\frac{2J}{N_s} \sum_{{\bf k},\omega_n} \frac{\Phi \gamma^2({\bf k})}
{(i\omega_n-\epsilon_{\bf k})(i\omega_n-\epsilon_{\bf k-Q})-[\Phi \gamma({\bf k})]^2} \;.\nonumber\\
\end{eqnarray}

In Fig. \ref{Tflux}(a) solid line shows the temperature $T_{FP}$ as a function of doping where the order parameter 
$\Phi$ is developed, i.e., where the FP instability takes place. 
$T_{FP}$ decreases when doping increases, and vanishes at a QCP at $\delta_c$. At $T=0$ the order parameter $\Phi$ decreases with doping as 
depicted by the solid line in Fig. \ref{Tflux}(b). In Fig. \ref{Tflux}(c) we show, for $\delta=0.10$, the
behavior of $\Phi$ as a function of the temperature. 
We note here that the $t$-$J$ model treated in large-N offers a route for describing the
$d$-CDW phase, where $\Phi$ and $T_{FP}$ are obtained  from the microscopic treatment and   
not introduced as input parameters. 

\subsection*{Self-energy fluctuations above $T_{FP}$}

\begin{figure}
\begin{center}
\setlength{\unitlength}{1cm}
\includegraphics[width=8cm,angle=0.]{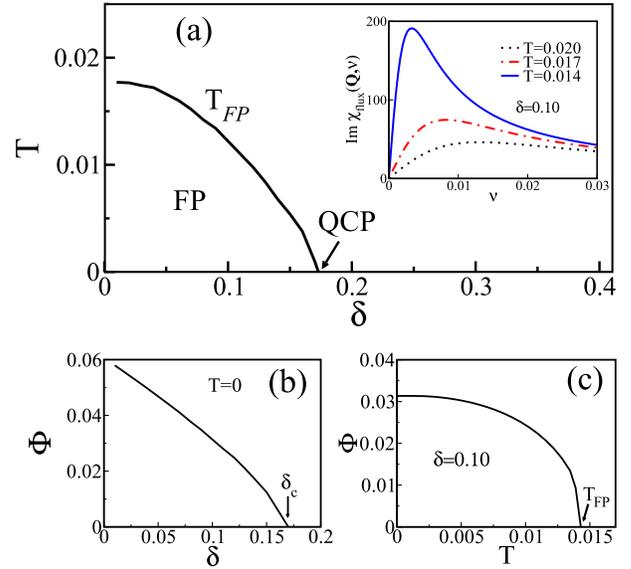}
\end{center}
\caption{(Color online) (a) $T_{FP}$ as a function of doping. At $T_{FP}$ the Fermi liquid state becomes unstable 
against a flux phase or $d$-CDW. The inset shows the softening of the $d$-wave flux mode approaching $T_{FP}$ for $\delta=0.10$. 
    (b) Flux phase order parameter $\Phi$ as a function of doping for $T=0$.  (c) $\Phi$ versus temperature for $\delta=0.10$. 
        $\eta=0.02$ was used in the analytical continuation $i\nu_n \rightarrow \nu+i\eta$.
}
\label{Tflux}
\end{figure}

Since $D_{ab}$ enters explicitly in the self-energy expression [Eq.(\ref{SigmaIm})], 
$\Sigma({\bf k},\omega)$ probes the proximity to the flux instability. 
At dopings and temperatures near the FP instability, the most important contribution 
to the self-energy  can be obtained after projecting  $\Sigma(k,\omega)$
on the FP eigenvector \textbf{\textit{v}}. The obtained self-energy is\cite{greco08,greco09}
\begin{eqnarray}
\label{sigmaflux}
{Im} \, \Sigma_{flux}({\bf k},\omega) &=& 
-\frac{J^2}{N_{s}} \sum_{{\bf q}} \Gamma^2({\bf q},{\bf k})
{Im} \chi_{flux}({\bf q},\omega-\epsilon_{{{\bf k}-{\bf q}}}) \nonumber \\
&\times& \left[n_{F}(-\epsilon_{{{\bf k}-{\bf q}}}) + n_{B}(\omega-\epsilon_{{{\bf k}-{\bf q}}})\right]
\end{eqnarray}
which shows the explicit contribution of the flux susceptibility
\begin{eqnarray}
\chi_{flux}({\bf q},\nu)= [2J\;r^2-\Pi({\bf q},\nu)]^{-1}\;.
\end{eqnarray}
In the above expression $\Pi({\bf q},i\nu_n)$ is
\begin{eqnarray}
\Pi({\bf q}, i\nu_n) = - \frac{J^2}{N_{s}}\;
\sum_{{\bf k}}\; \Gamma^2({\bf q},{\bf k}) \frac{[n_{F}(\epsilon_{{\bf k} - {\bf q}}) 
- n_{F}(\epsilon_{{\bf k}})]} 
{i \nu_n+\epsilon_{{\bf k} - {\bf q}} - \epsilon_{{\bf k}}}\;, \nonumber \\ 
\end{eqnarray}

\noindent where $\Gamma({\bf q},{\bf k})=r [\sin(k_x-q_x/2)-\sin(k_y-q_y/2)]$. Since the 
instability takes place at ${\bf q}={\bf Q}$, $\Gamma({\bf Q},{\bf k})$ is then proportional to $\gamma({\bf k})$, 
as expected due to the $d$-wave character of 
the FP instability. It can be shown that 
the onset temperature $T_{FP}$, where the flux instability takes place [Fig. \ref{Tflux} (a)], is the same 
temperature where the static ($i\nu_n=0$) flux
susceptibility $\chi_{flux}$ diverges at ${\bf q}={\bf Q}$.\cite{greco09,greco08}
At $T\rightarrow T_{FP}$  a soft mode [$Im\, \chi_{flux}({\bf Q},\nu)$] reaches $\nu=0$ freezing
the FP.\cite{greco09}
Inset of Fig. \ref{Tflux}(a) shows the softening of this mode approaching $T_{FP}$ from above for $\delta=0.10$. 
As shown below, the mode associated with the FP instability plays an important role in $\Sigma_{flux}({\bf k},\omega)$
at low doping.

\section{Pseudogap self-energy effects: comparison with ARPES and ADMR experiments}\label{FAMF}

In the FP state, below $T_{FP}$, the one-particle Green function [i.e., element $(1,1)$ from Eq.(\ref{G0phi})] is
\begin{eqnarray}\label{eq:GFP}
G({\bf k},i\omega_n)=\frac{1}{i\omega_n-\epsilon_{\bf k} - \Sigma_{FP}({\bf k},\omega_n)}
\end{eqnarray}
\noindent where 

\begin{eqnarray}\label{selfFP}
\Sigma_{FP}({\bf k},i\omega_n) = \frac{[\Phi\gamma({\bf k})]^2}{i\omega_n -\epsilon_{\bf k-Q}}
\end{eqnarray}

At this point it is important to remark the difference between $\Sigma_{FP}$ [Eq.(\ref{selfFP})] and 
$\Sigma_{flux}$ [Eq.(\ref{sigmaflux})]. While the former is an effective self-energy that describes the broken 
symmetry state (see also Ref.[\onlinecite{norman07}]), the second one represents dynamical self-energy effects 
in the paramagnetic state. 

\begin{figure}
\begin{center}
\setlength{\unitlength}{1cm}
\includegraphics[width=7cm,angle=0.]{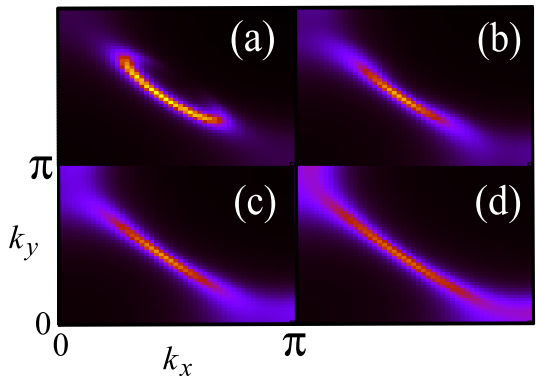}
\end{center}
\caption{(Color online) 
Short-ranged flux phase results. 
Panels (a)-(d) show results for the spectral intensity at $\omega=0$ vs $k_x$,$k_y$ for $\delta=0.07$
for $\xi \rightarrow \infty$, $\xi=5$, $\xi=2$ and $\xi=1$ respectively. For $\xi \rightarrow \infty$
[panel (a)] pockets predicted by the long-ranged FP case are obtained, while for the short-ranged case [panels (b)-(d)] 
the pockets wash out resembling an arc.
}
\label{ARPESshort}
\end{figure}

The Green function (\ref{eq:GFP}) leads to a FS  that consists of four hole pockets around nodal direction
with low spectral weight in the outer part of the pockets\cite{chakravarty03} [see Fig. \ref{ARPESshort}(a)].
However, most of the ARPES reports show disconnected arcs
\cite{damascelli03,norman98,kanigel06,kanigel07,shi,terashima07,yoshida09,kondo07,kondo09,lee07,
tanaka06,ma08,hashimoto10,ruihua2011} and to a lesser extent show pockets,\cite{chang08,razzoli10,meng09,yang08,yang09,yang11}
without full agreement on shape and localization of the pockets inside the Brillouin zone. 
Since experiments do not show clearly any turn at the end of the arc, the existence of a pocket
and the breaking of translational symmetry is doubtful.\cite{norman07} 
Furthermore, there are several drawbacks to be addressed. FP, as well as $d$-CDW,  predicts well defined QP 
peaks everywhere on the Brillouin zone while most of the ARPES show coherent 
peaks only near the nodal direction,  and near the antinode 
the QP peak is always broad. \cite{kim98,damascelli03}
On the other hand, in the FP  scenario a true phase
transition takes place at $T_{FP}$ where a true gap $\Phi$ is developed. 
However, experiments show a smooth crossover in the temperature behavior of normal state properties 
instead of an abrupt transition.\cite{kanigel06,kanigel07,tallon01} 
Consistent with this observation, ARPES shows that the PG is filling up but not closing with increasing
$T$, giving the impression of a smooth crossover of the spectral properties.\cite{norman98,kanigel06,kanigel07} 

These points can be discussed phenomenologically by assuming a short-ranged FP (or $d$-CDW) scenario.\cite{morinari09} 
In this context the self-energy can be written as

\begin{eqnarray} \label{Sigmashort}
\Sigma^{(SR)}_{FP}({\bf k},i\omega_n) = \sum_{\bf q} \frac{
P({\bf q}) [\Phi\gamma({\bf k-q})]^2}{i\omega_n-\epsilon_{{\bf k}-{\bf q}}}\;,
\end{eqnarray}
\noindent where $P({\bf q})$ represents a Lorentzian distribution of ordering 
wave  vector ${\bf Q}$  with a width $1/\xi$ associated 
with a finite correlation length $\xi$.

Using the obtained $\Phi$ of Fig. \ref{Tflux}(b), we show in Fig. \ref{ARPESshort}(a)-(d) the underlying FS  
for $\xi\rightarrow\infty$, $\xi=5$, $\xi=2$ and $\xi=1$  respectively, for $\delta=0.07$. 
While for $\xi\rightarrow\infty$ pockets are obtained, with decreasing $\xi$ the pockets wash out resembling an arc. In
addition, the length of the arcs increases with decreasing $\xi$. 

Although the short-ranged scenario seems to be better than the long-ranged case for describing ARPES, 
there are still several drawbacks:
(a) The doping and temperature dependence for the correlation length $\xi$ and $\Phi$ is required phenomenologically in order to fit 
the doping and temperature dependence of ARPES features.
(b) The presence of a finite $\Phi$ means also that a true phase transition exists 
at a certain temperature $T^*\equiv T_{FP}>T_c$, below which a true gap $\Phi$ emerges. As discussed above, this point 
is not clear from an experimental point of view. 
(c) The full Fermi surface is recovered either if $T>T_{FP}$ ($\Phi=0$) or when the correlation length is
lower than the lattice constant $a$ ($\xi\ll 1$). In both latter cases, the resulting spectral function is isotropic along
the FS.
As discussed above, ARPES experiments show that well defined QP peaks are observed only near the nodal direction, while 
near the antinodal direction the QP peaks are always broad,\cite{kim98,damascelli03}
i.e., the shape of the spectral function is very anisotropic on the FS even for $T>T^*$. 
(d) Finally, the $\omega$ dependence of $\Sigma^{(SR)}_{FP}$ is inherited only from the $\omega$
dependence of the bare Green function [Eq.(\ref{Sigmashort})], so that dynamical properties as transport cannot be
discussed in this context.

Next we will show that the presence of 
$\Sigma_{flux}$ in the paramagnetic state near the instability accounts for several observed features, 
without assuming additional phenomenological parameters or approximations beyond 
the large-N approximation.

\begin{figure}
\begin{center}
\setlength{\unitlength}{1cm}
\includegraphics[width=8.5cm,angle=0.]{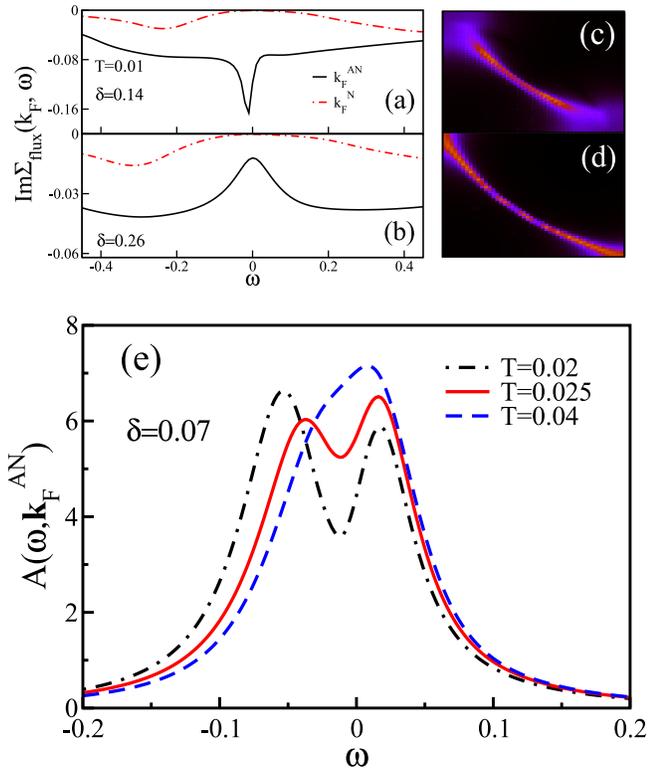}
\end{center}
\caption{(Color online) 
(a) and (b) imaginary part of the self-energy for the nodal (dotted dashed line)
and the antinodal (solid line) ${\bf k_F}$ for doping $\delta=0.14$ and
$\delta=0.26$ respectively, at $T=0.01$. (c) and (d) underlying FS
for $\delta=0.14$ and $\delta=0.26$ respectively. While for $\delta=0.26$
the expected large hole FS is obtained, for $\delta=0.14$ the unusual dips
near the antinode presented by the self-energy [panel (a)] develops an arc. (e) spectral
function obtained with $\Sigma_{flux}$ at $\delta=0.07$ for several temperatures. A PG-like feature
is developed at temperatures near but above $T_{FP}$, and fills with increasing temperature, 
in agreement with that observed in ARPES experiments.
}
\label{fluxarcs}
\end{figure}

\begin{figure}
\begin{center}
\setlength{\unitlength}{1cm}
\includegraphics[width=8.5cm,angle=0.]{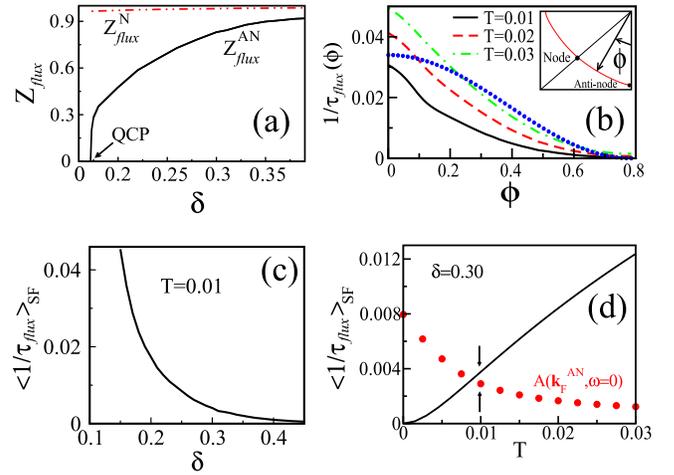}
\end{center}
\caption{(Color online)
(a) QP weight $Z_{flux}$ at $T=0$ as a function of doping, for the
nodal ($Z^{N}_{flux}$) and antinodal ($Z^{AN}_{flux}$) momentum on the FS. The QP weight is strongly
anisotropic on the FS, and vanishes at the antinode upon approaching the QCP.
(b) anisotropic scattering rate $1/\tau_{flux}({\bf k_F})=-2Im \Sigma_{flux}({\bf k_F},\omega=0) $,
on the FS for $\delta=0.20$ and several temperatures
as a function of the FS angle $\phi$ running from the
antinode $\phi=0$ to the node $\phi=\pi/4$ (shown in the inset).
$1/\tau_{flux}$ follows approximately a $d$-wave shape (dotted line).
(c) average on the FS of $1/\tau_{flux}$ at $T=0.01$ as a function of doping, showing that $\langle1/\tau_{flux}\rangle_{FS}$ 
is strongly doping dependent and vanishes in OD.
(d) average on the FS of $1/\tau_{flux}$ (solid line) and the corresponding spectral function intensity (dots) at ${\bf k^{AN}_F}$ and $\omega=0$, 
for $\delta=0.30$ as a function of temperature. 
Below a certain temperature (black arrows), the spectral function intensity at $\omega=0$ increases with decreasing $T$ faster than above
that temperature. 
The figure shows that
this temperature is close to that where a change in the curvature of $\langle1/\tau_{flux}\rangle_{FS}$ vs $T$ occurs.
}
\label{ADMRflux}
\end{figure}

As discussed in Sec. \ref{BG_A}, upon approaching $T_{FP}$ from above 
$Im \,\chi_{flux}({\bf Q},\omega)$ develops a soft energy mode [inset of Fig. \ref{Tflux}(a)] at
low doping and near $T_{FP}$.
Thus, contributions from $\Sigma_{flux}$ are the leading ones at low
doping and low energy. In addition, they are strongly dependent on $J$ and
doping,\cite{foussats04} and have $d$-wave anisotropy.
Figure \ref{fluxarcs} shows $Im \Sigma_{flux}$ at the Fermi momenta ${\bf k_F^N}$ (nodal) and ${\bf k_F^{AN}}$ (antinodal)
for $T=0.01$, for $\delta=0.14$ [panel (a)] and $\delta=0.26$ [panel (b)].
At ${\bf k_F^N}$ $Im\Sigma_{flux}$ is weak and  
leads to a well defined QP peak in the nodal direction for all doping.\cite{bejas11,greco09,greco08}
However, the behavior at ${\bf k_F^{AN}}$ is very peculiar at low doping.
Instead of a maximum at $\omega = 0$, $Im\Sigma_{flux}$ shows a minimum in the form of a dip. 
This behavior, which is in contrast to the expected one for a Fermi-liquid,\cite{katanin03,dellanna06} 
is the main reason for the PG [Fig. \ref{fluxarcs}(e)] and Fermi arc (FA) formation [Fig. \ref{fluxarcs}(c) for $\delta=0.14$]. 
With increasing doping and temperature the dip at $\omega \sim 0$ washes out, 
and $Im \,\Sigma_{flux}$ develops the expected maximum at $\omega = 0$ [Fig. \ref{fluxarcs}(b)], 
leading to the expected large hole FS [Fig. \ref{fluxarcs}(d) for $\delta=0.26$]. 
In addition, the leading edge of the PG near the antinode fills up
with temperature as in the experiments [Fig. \ref{fluxarcs}(e)].\cite{greco09,greco08,bejas11} 
From the above discussion, we use also the name PG self-energy for 
$\Sigma_{flux}$. As already discussed in Ref.[\onlinecite{greco11}], our approach offers a microscopic theory for understanding the 
phenomenological short-ranged order approach and may describe 
recent ARPES experiments.\cite{hashimoto10}

In Fig. \ref{ADMRflux}(a) we show the QP weight $Z_{flux}=(1-\frac{\partial Re\Sigma_{flux}}{\partial \omega}| _{\omega=0})^{-1}$
at $T=0$ for the FS points at the node (double dotted dashed line) and the antinode (solid line). The difference between $Z_{flux}^{AN}$
and $Z_{flux}^{N}$ shows the large anisotropy of $\Sigma_{flux}$ on the FS. Besides, $Z_{flux}^{AN}$ vanishes at the QCP.\cite{greco08}

In contrast to $\Sigma^{(SR)}_{FP}$, $\Sigma_{flux}$ can be used for discussing transport.
$\Sigma_{flux}$ leads to an anisotropic scattering rate on the FS, i.e.,  
$1/\tau_{flux}({\bf k_F}) \equiv -2 Im\, \Sigma_{flux}({\bf k_F},\omega=0)$.
As recently suggested,\cite{buzon10} $1/\tau_{flux}({\bf k_F})$ possesses an 
anisotropy on the FS close to $d$-wave [Fig. \ref{ADMRflux} panel (b)], and vanishes in OD [panel (c) for
the average in the FS $\langle1/\tau_{flux}\rangle_{FS}$]. 
As discussed in ADMR experiments these are the main characteristics 
for one of the two observed scattering rates.\cite{abdel06,abdel07,french09}

\begin{figure}
\begin{center}
\setlength{\unitlength}{1cm}
\includegraphics[width=8cm,angle=0.]{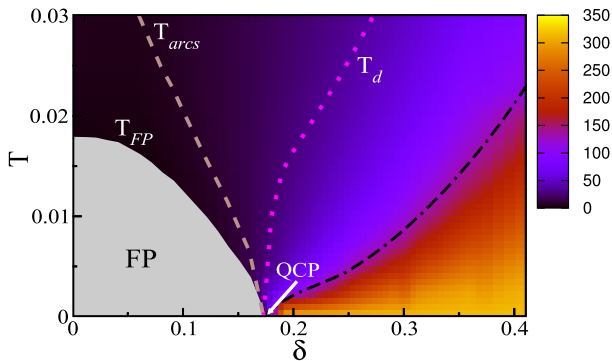}
\end{center}
\caption{(Color online) 
 Temperature versus doping phase diagram without superconductivity. 
In the gray area below $T_{FP}$ the translational symmetry is broken and pockets are obtained.
$T_{arcs}$ (dashed line) shows the crossover temperature below which the spectral functions have a pseudogap near the antinode
and Fermi arcs are developed. Therefore, we can associate $T_{arcs}$ with $T^*$.
Dotted line ($T_d$) is the temperature above which the self-energy presents a dip in $\omega=0$.
The background color shows the antinodal spectral function intensity at $\omega=0$. 
Double dashed dotted line shows the crossover from a bad metal at
low doping and high temperature to a good metal at high doping and low
temperature.
}
\label{Tcvsdp1}
\end{figure}

The background color in Fig. \ref{Tcvsdp1} shows, in the $T$-$\delta$ plane, the spectral function intensity obtained with
$\Sigma_{flux}$, at $\omega=0$ for the antinodal ${\bf k_F}$. This figure shows a crossover (double dashed dotted line) 
from a region of low intensity at low doping and high temperature
to a region of high intensity at large doping and low temperature.\cite{greco08}
As in Ref. [\onlinecite{kaminski03}] we identify the former region with an incoherent
metal where the QP peak is broad, whereas in the latter region the
QP is well defined as expected for a coherent metal. Note that the coherent to incoherent crossover, 
and its relation with the PG, can not be discussed in the context of the phenomenological short-ranged scenario.

Similar to Ref. [\onlinecite{kaminski03}],
above the crossover temperature (which increases with doping) the intensity of the
spectral function decreases with $T$ slower than below the crossover temperature. This crossover is also linked to the
behavior of the anisotropic scattering rate. In Fig. \ref{ADMRflux}(d) 
we show with black arrows that the spectral intensity (dots) increases faster with decreasing
temperature at approximately the same temperature where a change in the curvature of $\langle1/\tau_{flux}\rangle_{FS}$ (solid line) occurs. 
In Refs.[\onlinecite{kaminski03,castro04}] 
this crossover was also identified with a change in the temperature behavior of the resistivity. 
In Ref. [\onlinecite{varma97}] the incoherent-coherent crossover
was also discussed in terms of the paramagnetic state and the QCP scenario.

Dashed line in Fig. \ref{Tcvsdp1} ($T_{arcs}$) shows a crossover temperature above which the large hole FS is recovered. Thus, 
FAs occur between $T_{FP}$ and $T_{arcs}$ in the paramagnetic phase and without the breaking of translational
symmetry. Although dashed line ends at the QCP, $T_{arcs}$ is not a critical transition temperature.
Interestingly, $T_{arcs}$ is close to $T^{up}$ obtained in reference [\onlinecite{buzon10}] which is defined as
the temperature where $\langle1/\tau_{flux}\rangle_{FS}$ shows an upturn, allowing us to relate the onset of 
the pseudogap phase observed in ARPES with that obtained by resistivity experiments.\cite{daou09}
Thus, we associate $T_{arcs}$ with $T^*$.

Although arcs wash out above $T_{arcs}$, 
the self-energy does not behaves as expected from a Fermi liquid (FL). Between $T_{arcs}$ and $T_d$ (dotted line),
$\Sigma_{flux}$ shows a dip near the antinode, and only below $T_d$ 
the expected maximum at $\omega=0$ for the whole FS is obtained. The fact that there are no arcs 
between $T_{arcs}$ and $T_d$ can be explained as follows. For the existence of arcs $\Sigma_{flux}$ must show a dip at $\omega\sim0$ 
[see Fig. \ref{fluxarcs}(a)], but in addition, the height of the dip must be larger than its half-width. 
See Ref.[\onlinecite{yamase12}] where this point was discussed in the context of 
the Pomeranchuk instability where broad spectral functions on the FS and no PG were obtained, but the
self-energy is not FL-like. Thus, between $T_{arcs}$ and $T_d$ while the full FS is recovered, the shape of the spectral
functions is very anisotropic on the FS, i.e., QP peaks are broad near the antinode and sharp near the node.
Such anisotropy is lost with further increasing doping.

In summary, the fluctuating FP approach has 
several advantages over the short-ranged case:
(a) The $T_{FP}$ line is defined from microscopic treatment 
and depends only on the expression of $\chi_{flux}$, which
on the other hand enters in $\Sigma_{flux}$. In other words,
in the fluctuating PG framework there are no additional assumptions or input parameters beyond our
microscopic treatment of the $t$-$J$ model.
(b) There is no need for the occurrence of a true phase transition, since arcs are
originated dynamically due to the
interaction between carriers and FP fluctuations at $T>T_{FP}$, implying that long-range order is not broken.
(c) In the region between $T_{arcs}$ and $T_d$, although the full FS is recovered, it can be shown (see 
Fig. 12(b) of Ref.[\onlinecite{bejas11}]) that the spectral function shows well defined QP peaks
near the nodal direction and broad QP peaks near the antinodal direction, as observed in ARPES experiments.
(d) $\Sigma_{flux}$ 
contains its own  frequency and temperature dependence
which, as mentioned above, allows to discuss transport. 
Finally, we remark that for a more quantitative agreement with the phase diagram
of cuprates, $T_{FP}$ should be located at lower temperatures. See next section
for further discussion.

Cluster dynamical mean-field theory in Ref.[\onlinecite{lin10}] shows that 
the spectral functions develop a PG and FAs which change with doping and temperature in close 
agreement with our results. In addition, the PG and FAs formation were assigned to a self-energy behavior 
in which the imaginary part near $\omega=0$ shows a dip instead of a maximum. 
The agreement between our results and cluster dynamical 
mean-field theory is satisfactory. Although our approach is not exact, it offers a microscopic mechanism which
usually is less evident from purely numerical methods. 

In the next section we discuss the interplay between the PG and superconductivity.

\section{Superconductivity}\label{BG_B}

The large-N treatment of the $t$-$J$ model shows a
$d$-wave superconductivity which is mainly mediated by non-retarded interactions 
$J({\bf k}-{\bf k'})=2J [\cos(k_x-k'_x)+ \cos(k_y-k'_y)$].\cite{cappelluti99,zeyher98}
The gap equation for the $d$-wave superconducting order parameter $\Delta({\bf k})$ can be
written as:

\begin{eqnarray}\label{gap0}
\Delta({\bf k})=-4 J \gamma({\bf k}) \frac{T}{N_s} \sum_{{\bf k'},n'} \gamma({\bf k'})
G^0_{12}({\bf k'},i\omega_{n'})\;,
\end{eqnarray}

\noindent where $G^0_{12}$ is the element $(1,2)$ of the bare Green function 

\begin{equation} \label{G0}
(\widehat{G}^{0})^{-1}({\bf k},i\omega_{n}) = 
\left(
  \begin{array}{cccccc}
 i\omega_n-\epsilon_{\bf k}  &-\Delta({\bf k}) \\
 -\Delta({\bf k})     &i\omega_n+\epsilon_{\bf k}\\
\end{array}
\right).
\end{equation}

%

\begin{figure}
\begin{center}
\setlength{\unitlength}{1cm}
\includegraphics[width=9cm,angle=0.]{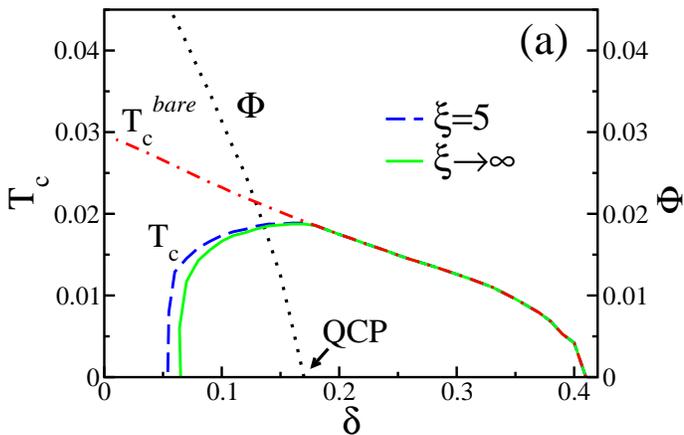}
\end{center}
\caption{(Color online)
Results of short-ranged FP  for $T_c$. Dotted line shows the gap $\Phi$ versus
doping [same as Fig. \ref{Tflux}(b)]. Dotted dashed line shows $T^{bare}_c$ obtained from 
Eq.(\ref{gap0}) without self-energies effects.
Solid line and dashed line are the results for $T_c$ obtained assuming in $\Sigma^{(SR)}_{FP}$ [Eq.(\ref{Sigmashort})] a
correlation length $\xi\rightarrow \infty$ and $\xi=5$, respectively. For $\delta>\delta_c$, $\Phi=0$, and therefore $\Sigma^{(SR)}_{FP}=0$, thus $T_c=T^{bare}_c$.
}
\label{Tcvsdpficorr}
\end{figure}

Figure \ref{Tcvsdpficorr} shows that the bare superconducting critical temperature, called $T^{bare}_c$  (dotted dashed line), 
increases monotonically with decreasing doping. The term bare is used here because the gap equation (\ref{gap0}) 
does not contain self-energy effects. We note that the mean-field superconducting gap equation (\ref{gap0}) is identical
to Eq. (\ref{phi}) for the FP order parameter $\Phi$ if $\epsilon_{{\bf k}-{\bf Q}}$
is replaced by $-\epsilon_{\bf k}$ in (\ref{phi}).

Taking into account self-energy effects, the gap equation results:

\begin{eqnarray}\label{gap}
\Delta({\bf k})=-4 J \gamma({\bf k}) \frac{T}{N_s} \sum_{{\bf k'},n'} \gamma({\bf k'}) G_{12}({\bf k'},i\omega_{n'})\;.
\end{eqnarray}

In Eq.(\ref{gap}) the renormalized Green function $\widehat{G}$ is:\cite{rickayzen}

\begin{eqnarray} \label{G}
\widehat{G}^{-1}({\bf k},i\omega_{n}) = (\widehat{G}^0)^{-1}({\bf k},i\omega_{n})-\widehat{\Sigma}({\bf k},i\omega_n)
\end{eqnarray}
\noindent where 
\begin{eqnarray}\label{eq:Sigmagen}
\widehat{\Sigma}({\bf k},i\omega_n)=
\left(
  \begin{array}{cccccc}
 \Sigma({\bf k},i\omega_n)  &0 \\
 0 &-\Sigma({\bf k},-i\omega_n)\\
\end{array}
\right).&
\end{eqnarray}

Eq.(\ref{gap}) means that we assume that excitations entering in $\Sigma$ do not affect the non-retarded pairing.

\subsection*{Short-ranged flux phase case}

Figure \ref{Tcvsdpficorr} shows results for  $T_c$ obtained using the self-energy $\Sigma^{(SR)}_{FP}$ (\ref{Sigmashort}) in (\ref{eq:Sigmagen}). 
For the long-ranged case $\xi\rightarrow\infty$ (solid line), results are in close agreement with the
treatment of Ref.[\onlinecite{cappelluti99}], where the competition between $\Phi$ and
$\Delta$ was considered self-consistently. 
Interestingly, even for short correlation length of about 
$\xi=5$, we obtain a well defined DSB for $T_c$ (dashed line). Since 
for $\delta > \delta_c$, $\Phi=0$ ($\Sigma^{(SR)}_{FP}=0$), then $T_c$ follows $T^{bare}_c$ (dotted dashed line). 

However, as discussed in the previous section, the FP fluctuating 
approach has several advantages over the short-ranged $d$-CDW. Thus, 
it is interesting to see if fluctuations near the FP instability can lead to a DSB for $T_c$.

\subsection*{Role of the fluctuating pseudogap self-energy $\Sigma_{flux}$ on superconductivity}\label{flux}

The role of the PG self-energy on the doping behavior of $T_c$ can 
be studied by the gap equation (\ref{gap}) considering $\Sigma_{flux}$ [Eq.(\ref{sigmaflux})]. 
In our view the PG fluctuations do not glue the electrons into Cooper pairs.

Since we are  interested in the doping behavior of $T_c$, it is not necessary to consider the effects 
of the superconducting gap on 
$\Sigma_{flux}$ which, then, can be calculated in the normal state.

We remark that $d$-wave superconductivity\cite{dagotto93} 
and a DSB\cite{kancharla08,sorella02} for $T_c$ in 
the $t$-$J$ model were supported by numerical studies.  

\begin{figure}
\begin{center}
\setlength{\unitlength}{1cm}
\includegraphics[width=8cm,angle=0.]{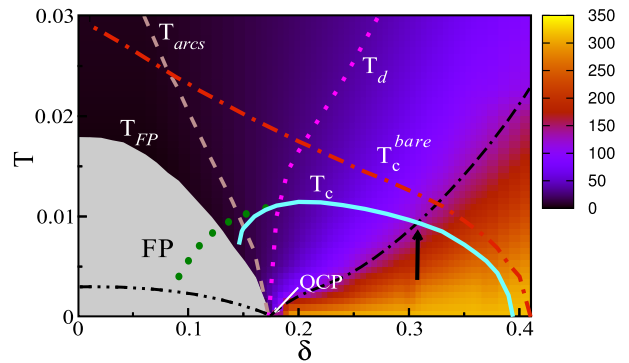}
\end{center}
\caption{(Color online) 
Temperature versus doping phase diagram with the inclusion of superconductivity. 
Dotted dashed, dashed and dotted line shows $T^{bare}_c$ (same as Fig.\ref{Tcvsdpficorr}), $T_{arcs}$ and $T_d$ respectively. 
Solid line shows the dome-shaped behavior of 
$T_c$ obtained in presence of $\Sigma_{flux}$. Interestingly, $T_c$ go first
parallel to $T^{bare}_c$, and shows a downward deviation (black arrow) at the crossover temperature from a coherent to
an incoherent metal. Double dotted dashed line illustrates a situation where the FP instability occurs at lower temperature.
The expected superconducting temperature for this situation is illustrated by the dots next to $T_c$ (see text for discussion).
}
\label{Tcvsdp3}
\end{figure}

Figure \ref{Tcvsdp3} shows that $T_c$ (solid line) moves at first parallel to the results for $T^{bare}_c$
(dotted dashed line) with decreasing doping. For high doping and low temperature,
$\Sigma_{flux}$ behaves as in a conventional Fermi-liquid, presenting a maximum at $\omega=0$ as discussed in the previous section.
Then, for $\delta\gtrsim0.3$ the reduction of $T_c$ from $T^{bare}_c$ is due to conventional Fermi-liquid 
self-energy effects where $T_c$ diminishes following approximately the 
form $\log\,(\,T_c/\,T_c^{bare}) \sim (1-Z^{-1})$.

However, $T_c$ does not follow the same trend as $T_{c}^{bare}$ for $\delta<0.30$. 
$T_c$ deviates from that trend at about the crossover line between the coherent and
the incoherent metal (see black arrow in Fig. \ref{Tcvsdp3}).
Approaching the  QCP, $T_c$ reaches its maximum and begins to fall with decreasing doping, showing 
clear tendencies to develop a DSB.  
The fact that the fall of $T_c$ occurs approaching the  QCP allows to
associate this behavior to the opening of the fluctuating PG, showing the relevance 
of critical fluctuations near the instability. 
This fact shows that the DSB of $T_c$ is related to the properties of the
incoherent metal near the instability, and not to the coherent metal
at low temperature and large doping. 
It is important also to remark that the coherent metal properties predict that $T_c$ 
would vanishes at the QCP where $Z \rightarrow 0$ [Fig. \ref{ADMRflux}(a)]. However, $T_c$ tends to show a DSB for 
dopings lower than the QCP which indicates also the 
relevance of critical fluctuations in the behavior of $T_c$.  

In a recent cluster dynamical mean-field calculation\cite{gull13} it was also discussed that the maximum of the superconducting 
transition temperature occurs near the onset of the PG.

Note that we stop the calculation of $T_c$ at the onset of $T_{FP}$, i.e., we do not enter inside the FP instability.
For a more quantitative agreement with the phenomenology $T_{FP}$ should be at lower temperatures, e.g. following
double dotted dashed line sketched in Fig. \ref{Tcvsdp3}.
Under this condition the expected trend for the superconducting critical temperature is depicted by dots.
It is worth to mention that $T_{FP}$ is a mean-field critical temperature and should be interpreted
as a temperature scale where the corresponding charge order, at least its fluctuation effect, may become important.
It is known that the effect of fluctuations may considerably reduce the transition temperature with respect to the 
mean-field temperature.\cite{lee73}

In summary, the pseudogap fluctuations in the proximity to $T_{FP}$ 
may lead to a DSB for $T_c$ and, in addition, develop arcs in the paramagnetic phase which 
change with doping and temperature as seen in the experiments. 
In other words, for the existence of
FAs in the normal state it is not necessary that $T<T_{FP}$ because arcs are generated by fluctuations in the proximity 
to the FP instability and, that fluctuations seem to be enough for triggering a DSB for $T_c$. 
The phase diagram of Fig. \ref{Tcvsdp3} shows close analogies with the phase diagram observed in hole doped cuprates
(see for example Fig. 5 of Ref.[\onlinecite{kaminski03}] and Fig. 1 of Ref.[\onlinecite{castro04}]). Indeed, it has
a crossover between a coherent metal and an incoherent metal, a temperature $T^{arcs}$ which can be associated with 
the PG temperature $T^*$, and tendencies to a DSB for the critical temperature $T_c$.

\section{Discussion and conclusion}\label{Conclusion}

In the context of the $t$-$J$ model we have discussed the dome-shaped behavior of $T_c$ and  
its relation with the  pseudogap and normal state properties. 
The $t$-$J$ model shows non-retarded interactions $J({\bf k}-{\bf k'})$ which lead to $d$-wave 
superconductivity where $T_c$ increases monotonically with decreasing doping. 
Differently to conventional superconductivity, where pairing excitations
can also be seen in the normal state self-energy effects, non-retarded
pairing does not contribute to transport or other normal state property. However, self-energy effects may affect the doping behavior of $T_c$.  
We have shown that fluctuations near the flux phase (or $d$-CDW) instability, existing at low doping and low temperature in
the $t$-$J$ model, lead to self-energy effects 
which may show a dome-shaped behavior for $T_c$. Moreover, (a) these self-energy effects
may also describe the arc-physics observed in ARPES above $T_c$ without invoking a true phase transition and without 
translational symmetry breaking; (b) these self-energy effects 
are also supported by the existence of an anisotropic scattering rate 
observed in ADMR experiments. 

A final remark. There are several experimental indications about the existence of an isotropic self-energy
contribution with a high-energy scale, i.e., with properties different to those related to the PG. 
These experimental indications came from transport,\cite{cooper09} ADMR,\cite{abdel06,abdel07,french09} 
ARPES,\cite{xie07,meevasana07,graf07,zhang08,koitzsch04,bogdanov00,kordyuk06} optical conductivity,\cite{hwang07} and 
Raman.\cite{li12} Theoretical reports also support the existence of self-energy contributions with 
properties which are different to the PG.\cite{buzon10,foussats08,zemlic08,kokalj11}
If only a single universal physics is mainly involved in the DSB, the role of isotropic self-energy effects must be discussed, because
a strong doping dependence of this self-energy, mainly in UD, may mask the effects of the PG in the DSB for $T_c$.

\noindent{\bf Acknowledgments}

 The authors thank to P. Horsch, J. Kokalj, R.H. McKenzie, J. Riera, H. Yamase, and R. Zeyher for valuable discussions,
 and to H. Parent for critical reading of the manuscript.

\end{document}